\documentclass{aip-cp}

\usepackage[numbers,sort&compress]{natbib}
\usepackage{rotating}
\usepackage{graphicx}

\usepackage{graphicx}
\usepackage[english]{babel}
\usepackage{listings} 
\newcommand{\beq}{\begin{equation}}
\newcommand{\eeq}{\end{equation}}
\newcommand{\bea}{\begin{eqnarray}}
\newcommand{\eea}{\end{eqnarray}}
\newcommand{\real}{{\sf I}\kern-.12em{\sf R}}
\newcommand{\comp}{{\sf I}\kern-.50em{\sf C}}
\newcommand{\unity}{{\sf I}\kern-.54em{\sf 1}}

\newcommand{\stringa}{\ttfamily\lstinline}
\def\cod#1{{\stringa!#1!}}

\begin{document}

\title{Inclusive Three- and Four-jet Production 
       in Multi-Regge Kinematics at the LHC}

\author[aff1,aff2]{Francesco Caporale}
\eaddress{francesco.caporale@uam.es}
\author[aff1,aff2,aff3,aff4]{Francesco G. Celiberto \corref{cor1}}
\author[aff1,aff2]{Grigorios Chachamis}
\eaddress{chachamis@gmail.com}
\author[aff1,aff2]{D. Gordo G{\'o}mez}
\eaddress{david.gordo@csic.es}
\author[aff1,aff2]{Agust{\'i}n Sabio Vera}
\eaddress{a.sabio.vera@gmail.com}

\affil[aff1]{Instituto de F{\' \i}sica Te{\' o}rica UAM/CSIC, 
              Nicol{\'a}s Cabrera 15, 
              28049 Madrid, Spain}
\affil[aff2]{Universidad Aut{\' o}noma de Madrid, 
              28049 Madrid, Spain}
\affil[aff3]{Dipartimento di Fisica, Universit{\`a} della Calabria, 
              Arcavacata di Rende, 87036 Cosenza, Italy}
\affil[aff4]{Istituto Nazionale di Fisica Nucleare, 
              Gruppo Collegato di Cosenza, 
              Arcavacata di Rende, 87036 Cosenza, Italy}

\corresp[cor1]{Corresponding author: 
               francescogiovanni.celiberto@fis.unical.it}

\maketitle

\begin{abstract}
A study of differential cross sections for the production 
of three and four jets in multi-Regge kinematics is presented. 
The main focus lies on the azimuthal angle dependences 
in events with two forward/backward jets tagged in the final state. 
Furthermore, the tagging of one or two extra jets 
in more central regions of the detector 
with a relative separation in rapidity from each other is requested. 
It is found that the dependence of the cross sections 
on the transverse momenta and the rapidities 
of the central jet(s) can offer new means 
of studying the onset of BFKL dynamics.

\end{abstract}

\section{INTRODUCTION}

The study of semi-hard processes 
in the high-energy (Regge) limit
is an active research field in perturbative QCD, 
the Large Hadron Collider (LHC) 
affording an abundance of data. 
Multi-Regge kinematics (MRK),
which requires final-state objects 
strongly ordered in rapidity,
is the principal ingredient for the study 
of multi-jet production at LHC energies.
In this kinematical regime, 
the Balitsky-Fadin-Kuraev-Lipatov 
(BFKL) approach, at leading (LLA) 
\cite{Lipatov:1985uk,Balitsky:1978ic,
Kuraev:1977fs,Kuraev:1976ge,Lipatov:1976zz,Fadin:1975cb} 
and next-to-leading 
(NLA)~\cite{Fadin:1998py,Ciafaloni:1998gs} accuracy, 
represents the most powerful mechanism
to resum the large logarithms 
in the colliding energy 
which are present to all orders of the perturbative expansion.
This approach was successfully applied 
to Deep Inelastic Scattering at HERA 
(see, {\it e.g.}~\cite{Hentschinski:2012kr,
Hentschinski:2013id}) in order to study quite inclusive 
processes which, however, are not so suitable 
to discriminate between BFKL dynamics 
and other resummations. 
The high energies reachable at the LHC 
give us the opportunity to study reactions
with much more exclusive final states which can, in principle, 
be only described by the BFKL resummation, making it possible 
to unravel the applicability region of the approach. 
In the last years, 
Mueller--Navelet jet production~\cite{Mueller:1986ey} 
has been the most studied reaction.
Interesting observables associated to this process are 
the azimuthal correlation momenta which, however, 
seem to be strongly affected by collinear contaminations.
Therefore, new observables independent 
from the conformal contribution
were proposed in~\cite{Vera:2006un,Vera:2007kn} 
and calculated at NLA 
in~\cite{Ducloue:2013bva,Caporale:2014gpa,Caporale:2015uva,
Celiberto:2015yba,Celiberto:2016ygs,
Ciesielski:2014dfa,Angioni:2011wj,Chachamis:2015crx,N.Cartiglia:2015gve},  
showing a very good agreement 
with experimental data at the LHC. 
Unfortunately, Mueller-Navelet configurations 
are still too inclusive to accomplish MRK precision studies. 
With the aim to deeply probe the BFKL dynamics 
by studying azimuthal correlations where the transverse momenta
of extra particles introduce a new dependence, 
we define new observables for semi-hard processes 
which can be considered as a generalization 
of Mueller-Navelet jets\footnote{Another interesting 
and novel possibility, the detection of two charged light hadrons:
$\pi^{\pm}$, $K^{\pm}$, $p$, $\bar p$ 
having high transverse momenta and 
separated by a large interval of rapidity,
together with an undetected soft-gluon radiation emission, 
was suggested in~\cite{Ivanov:2012iv} 
and studied in~\cite{Celiberto:2016hae,Celiberto:2017ptm}.}. 
These processes are inclusive 
three-jet~\cite{Caporale:2015vya,Caporale:2016soq,Caporale:2016zkc}
and four-jet production~\cite{Caporale:2015int,Caporale:2016xku}.

\section{THREE- AND FOUR- JET PRODUCTION}

The kind of processes we want to study 
is the inclusive hadroproduction of $n$ jets in the final state, 
well separated in rapidity so that 
$y_i > y_{i+1}$ according to MRK, 
while their transverse momenta $\{k_i\}$  
lie above the experimental resolution scale,
together with an undetected gluon radiation emission.
Our goal is to generalize the azimuthal ratios $R_{nm}$ 
defined in the Mueller--Navelet jet configuration.
For this reason, we define new, generalized azimuthal observables 
by considering the projection of the differential cross section 
$d\sigma^{n-{\rm jet}}$
on all angles, so having the general formula 
given in Equation~(3) of~\cite{Celiberto:2016vhn} 
and in Equation~(1) of~\cite{Caporale:2016pqe}:
\begin{equation}
\mathcal{C}_{M_1 \cdots M_{n-1}} =
\left\langle 
 \prod_{i=1}^{n-1} \cos\left(M_i \, \phi_{i,i+1}\right)
\right\rangle = 
\int_0^{2\pi} d\theta_1 
\cdots  
\int_0^{2\pi} d\theta_n
\prod_{i=1}^{n-1} \cos\left(M_i \, \phi_{i,i+1}\right)
d\sigma^{n-{\rm jet}}\;,
\end{equation}
where $\phi_{i,i+1} = \theta_i - \theta_{i+1} - \pi$, 
and $\theta_i$ is the azimuthal angle of the $i$-th jet.

From a phenomenological point of view, 
our goal is to give predictions compatible 
with the current and future experimental data.
So, we introduce the kinematical cuts 
already used at the LHC by integrating 
$\mathcal{C}_{M_1 \cdots M_{n-1}}$ 
over the momenta of all tagged jets in the form
\begin{equation}\label{Cm_int}
C_{M_1 \cdots M_{n-1}} =
\int_{y_{1,\rm min}}^{y_{1,\rm max}}dy_1
\int_{y_{n,\rm min}}^{y_{n,\rm max}}dy_n
\int_{k_{1,\rm min}}^{k_{1,\rm max}}dk_1
\cdots
\int_{k_{n,\rm min}}^{k_{n,\rm max}}dk_n
\, \delta\left(y_1-y_n-Y\right)\,
\mathcal{C}_{M_1 \cdots M_{n-1}} \;,
\end{equation}
where the rapidities of the most forward 
and of the most backward jet lie in the range 
$y_1^{\rm min} = y_n^{\rm min} = -4.7$  and 
$y_1^{\rm max} = y_n^{\rm max} = 4.7$, keeping their difference 
$Y = y_1 - y_n$ fixed.
From a more theoretical aspect, 
it is crucial to improve the stability of our 
predictions (see~\cite{Caporale:2013uva} for a related discussion).
This can be carried out by removing 
the zeroth conformal spin contribution 
responsible for any collinear contamination.
Therefore, we introduce the ratios
\begin{equation}
R^{M_1 \cdots M_{n-1}}_{N_1 \cdots N_{n-1}} \equiv 
\frac{C_{M_1 \cdots M_{n-1}}}{C_{N_1 \cdots N_{n-1}}}\;,
\end{equation}
with $\{M_i\}$ and $\{N_i\}$ being positive integers. 

\begin{figure}[h]
\begin{minipage}{.5\textwidth}
\centering
\hspace{-0.7cm}
\includegraphics[scale=0.286]{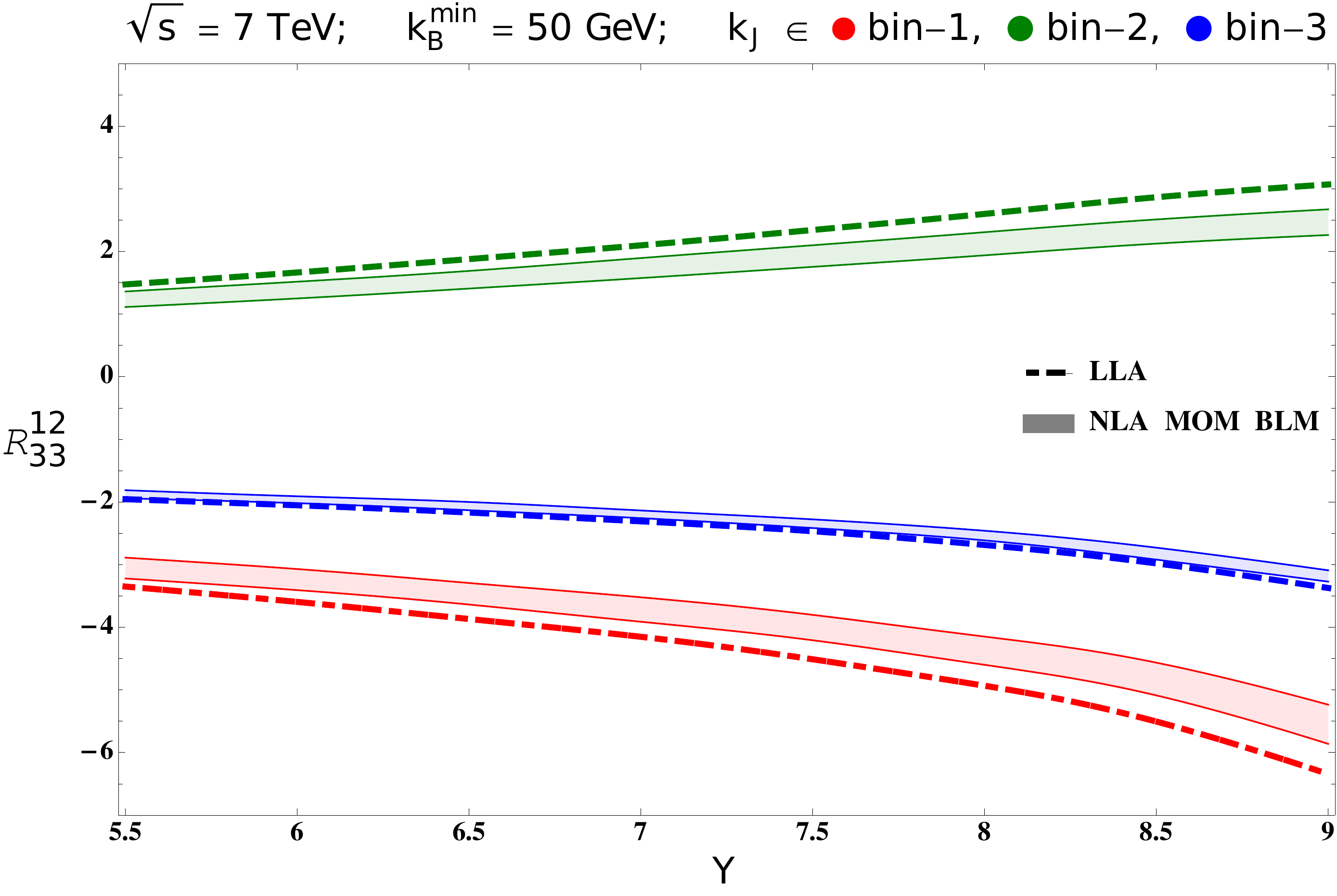}
\end{minipage}
\begin{minipage}{.5\textwidth}
\centering
\includegraphics[scale=0.286]{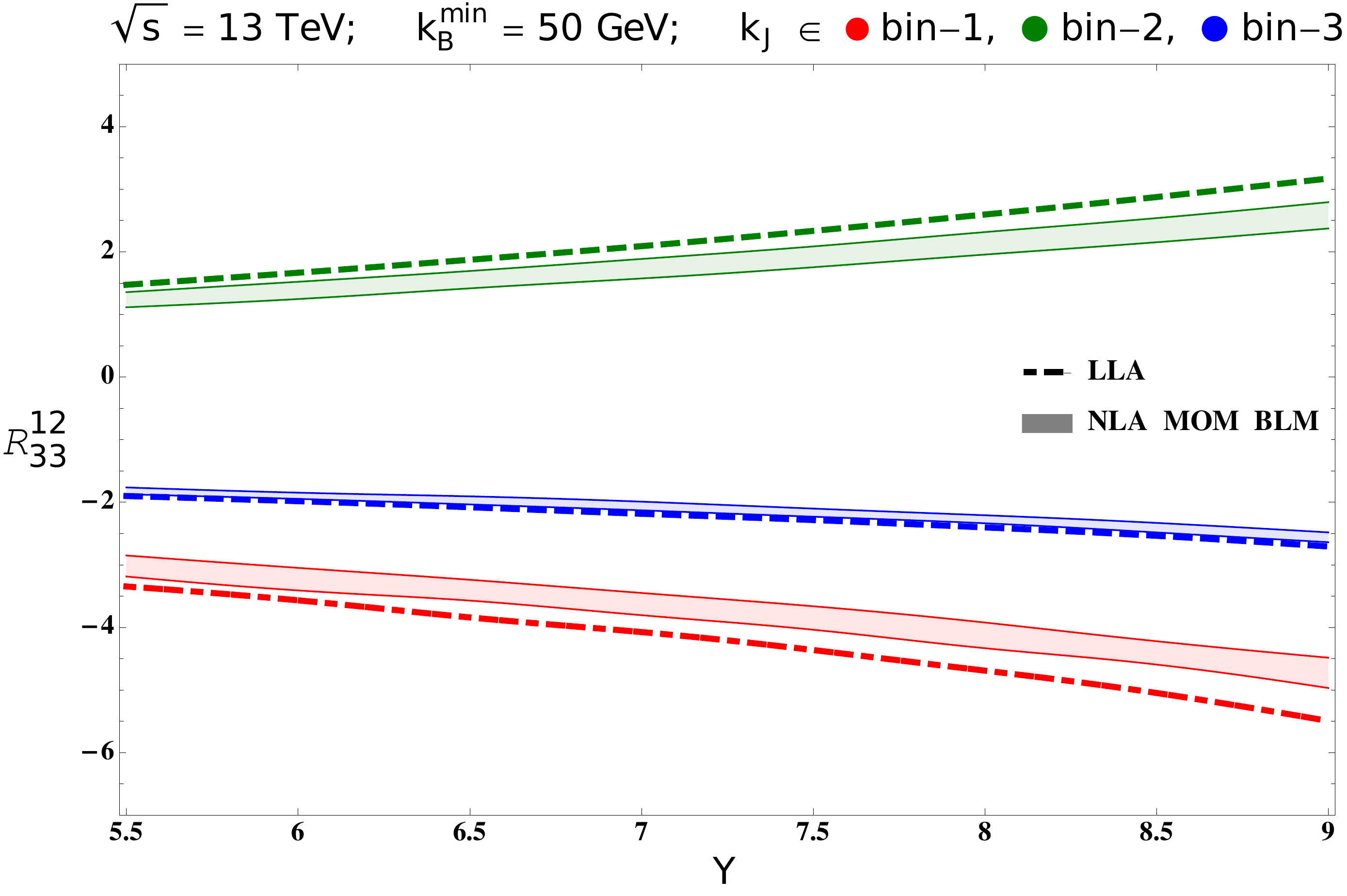}
\end{minipage}
\caption{$Y$-dependence of $R^{12}_{33}$ 
in the LLA and NLA 
(MOM BLM optimization method~\cite{Brodsky:1982gc} is used) accuracy 
for $\sqrt{s} = 7, 13$ TeV, 
$k_{A,\rm min} = 35$ GeV
$k_{B,\rm min} = 50$ GeV. 
The central jet rapidity is set to~$y_J = (y_A + y_B)/2$, 
while its transverse momentum~$k_J$ is allowed to take values 
in the following three ranges: 
[$20\, \mathrm{GeV} < k_J < 35\, \mathrm{GeV}$] 
(\emph{bin-1}),
[$35 \,\mathrm{GeV} < k_J < 60\, \mathrm{GeV}$] 
(\emph{bin-2}) and
[$60\, \mathrm{GeV} < k_J < 120\, \mathrm{GeV}$] 
(\emph{bin-3}).}
\label{fig:3jet}
\end{figure}

In Figure~\ref{fig:3jet} we present the dependence on $Y$ 
of the $R^{12}_{33}$ ratio, 
characteristic of the 3-jet process, 
for $\sqrt{s} = 7$ and $13$ TeV, for two different kinematical cuts 
on the most forward/backward jet transverse momenta $k_{A,B}$
and for three different ranges of the central jet
transverse momentum $k_J$, that is, 
$20\, \mathrm{GeV} < k_J < 35\, \mathrm{GeV}$ 
(\emph{bin-1}, smaller than $k_A$, $k_B$),
$35 \,\mathrm{GeV} < k_J < 60\, \mathrm{GeV}$ 
(\emph{bin-2}, similar to $k_A$, $k_B$) and
$60\, \mathrm{GeV} < k_J < 120\, \mathrm{GeV}$ 
(\emph{bin-3}, larger than $k_A$, $k_B$).  
We clearly see that the contribution of NLA corrections is small 
with respect to the LLA predictions.

\begin{figure}[h]
\begin{minipage}{.5\textwidth}
\centering
\hspace{-0.7cm}
\includegraphics[scale=0.229]{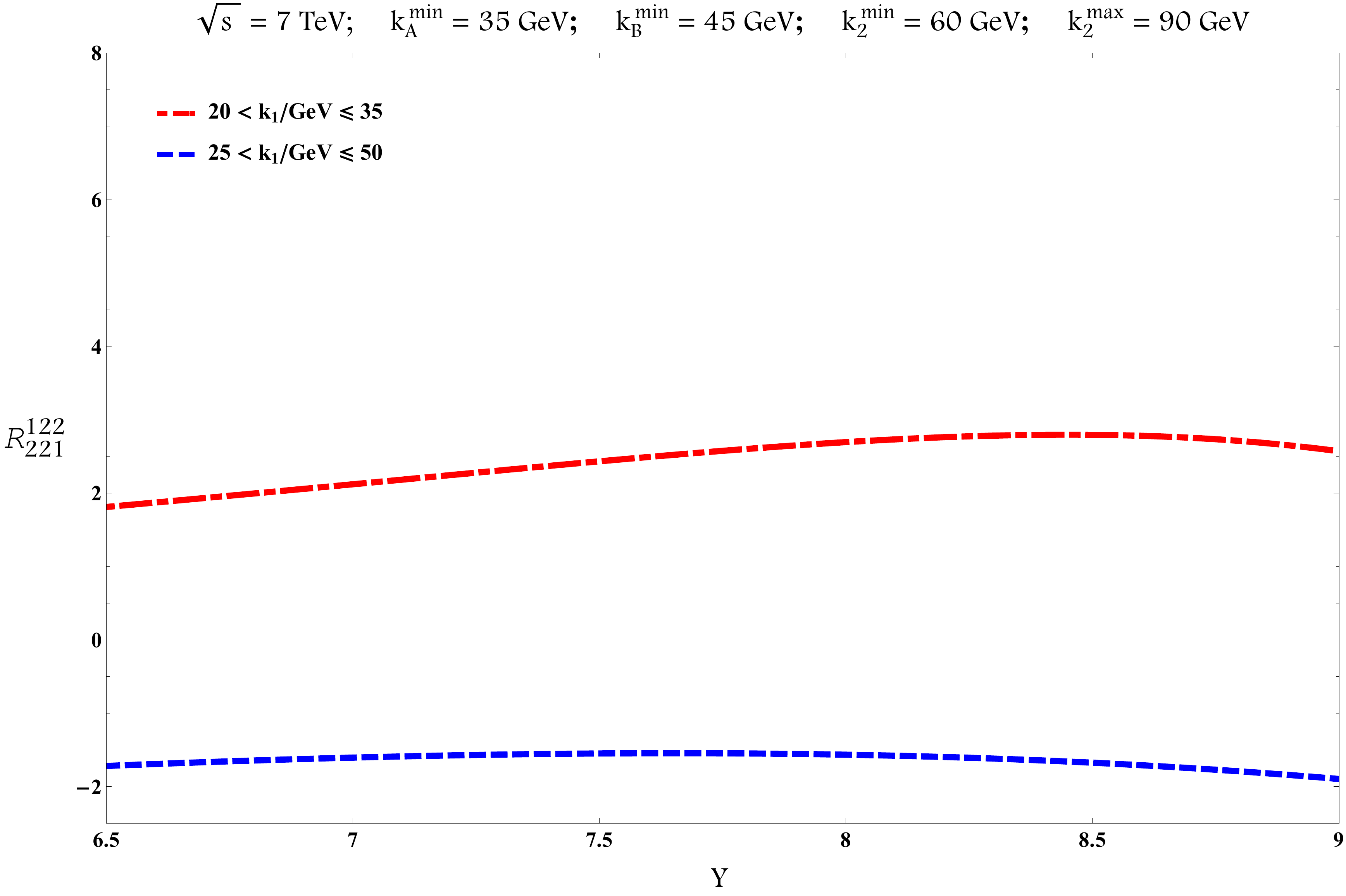}
\end{minipage}
\begin{minipage}{.5\textwidth}
\centering
\includegraphics[scale=0.229]{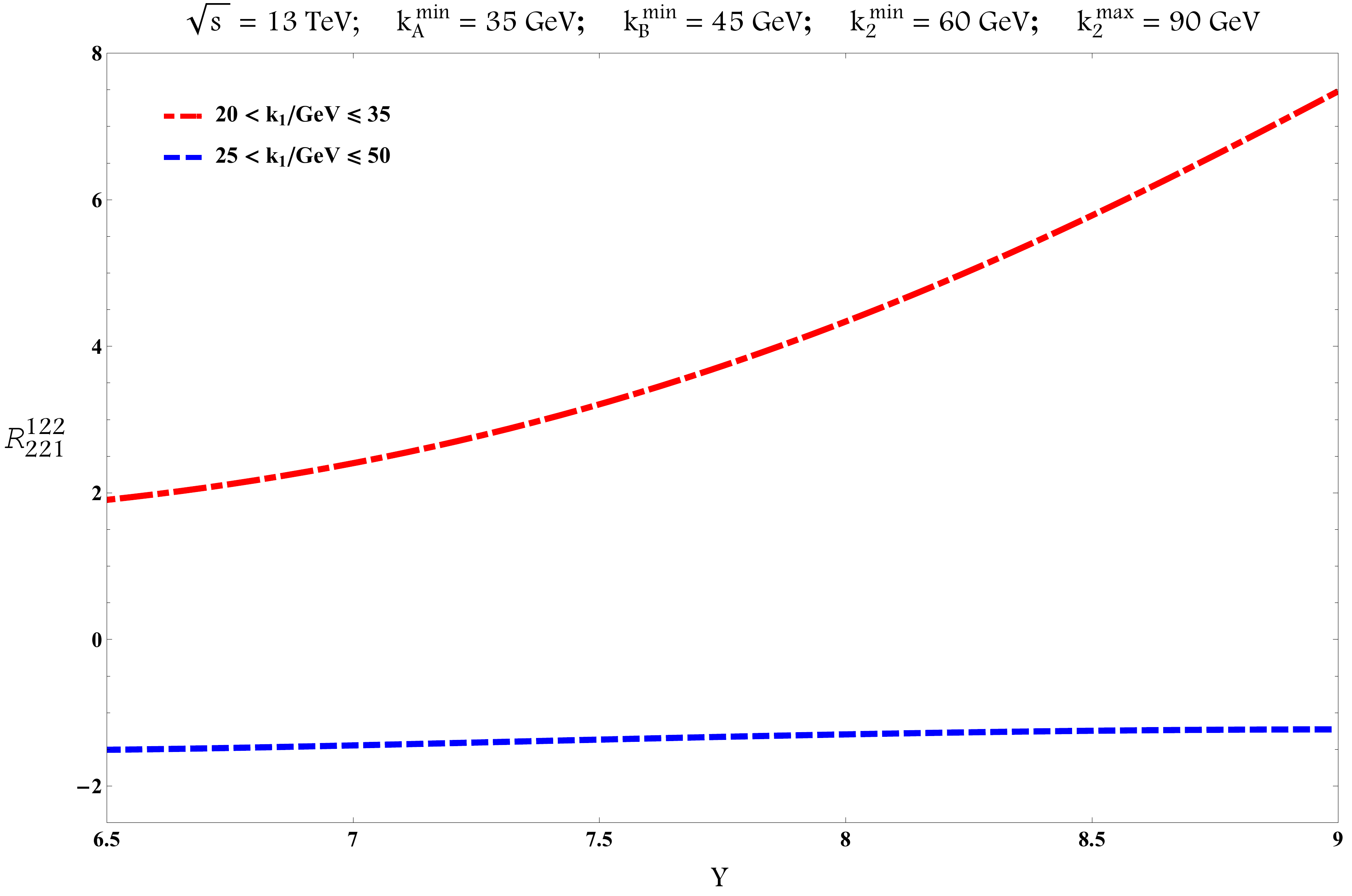}
\end{minipage}
\caption{$Y$-dependence of $R^{122}_{221}$
for $\sqrt{s} = 7$ TeV and for $\sqrt{s} = 13$ TeV. 
The rapidity interval between a jet and the closest one 
is fixed to~$Y/3$.}
\label{fig:4jet}
\end{figure}

In Figure~\ref{fig:4jet} we show the dependence on $Y$ 
of the $R^{122}_{221}$ ratio, 
characteristic of the 4-jet process, 
for $\sqrt{s} = 7$ and $13$ TeV, for asymmetrical cuts 
on the external jet transverse momenta $k_{A,B}$ 
and for two different configurations 
of the central jet transverse momenta $k_{1,2}$. 

A comparison with predictions for these observables from
fixed order analyses as well as from the BFKL inspired 
Monte Carlo {\bf\cod{BFKLex}}~\cite{Chachamis:2011rw,
Chachamis:2011nz,Chachamis:2012fk,
Chachamis:2012qw,Caporale:2013bva,Chachamis:2016ejm,
Chachamis:2015zzp,Chachamis:2015ico} 
is in progress.


\section{SUMMARY \& OUTLOOK}

We defined ratios of correlation functions of products 
of azimuthal angle difference cosines in order to study 
multi-jet production processes at hadron colliders,  
the dependence on the transverse momenta and rapidities 
of the central jet(s) being a distinct signal of the BFKL dynamics. 
In the case of three-jet production, we included 
the NLA contributions coming from the BFKL gluon Green function 
and used the MOM Brodsky-Lepage-Mackenzie 
(BLM) method~\cite{Brodsky:1982gc} 
in order to optimize the value of the renormalization scale $\mu_R$.
However, more accurate analyses are still needed: 
full NLA analyses including 
next-to-leading order jet vertices and study 
of different configurations for the rapidity range 
of the two central jets, together with the analysis of the effect 
of using different PDF parametrizations. 
It would be also interesting to study the behavior of 
our observables in other approaches 
not based on the BFKL resummation and to
test how they differ from the predictions given in this work.
Only experimental analyses of these observables using existing
and future LHC data will to probe and disentangle
the applicability region of the BFKL dynamics.
For this reason, we strongly suggest experimental collaborations 
to study these observables in the next LHC analyses.

\section{ACKNOWLEDGMENTS}

GC acknowledges support from the MICINN, Spain, 
under contract FPA2013-44773-P. DGG acknowledges 
financial support from `la Caixa'-Severo Ochoa doctoral fellowship.
ASV and DGG acknowledge support from the Spanish Government 
(MICINN (FPA2015-65480-P)) and, together with FC and FGC, 
to the Spanish MINECO Centro de Excelencia Severo Ochoa Programme (SEV-2012-0249). 
FGC thanks the Instituto de F{\'\i}sica Te{\'o}rica 
(IFT UAM-CSIC) in Madrid for warm hospitality.

\end{document}